\documentclass[preprint,aps]{revtex4}
\usepackage{graphicx}
\usepackage{amssymb}
\usepackage[a4paper]{geometry}
\usepackage{url}
\usepackage{setspace}

\begin{document}

\title{Soft core thermodynamics from self-consistent hard core fluids}

\author{Elisabeth Sch\"oll-Paschinger$^{1}$ and Albert Reiner$^{2}$}

\affiliation{ $^1$Fakult\"at f\"ur Physik, Universit\"at Wien,
  Boltzmanngasse 5, A-1090 Wien, Austria \\
  $^2$Teoretisk fysikk, Institutt for fysikk,\\
Norges teknisk-naturvitenskapelige universitet (NTNU) Trondheim,\\
H\o gskoleringen 5, N-7491 Trondheim, Norway}

\date{\today}

\begin{abstract}%%
  In an effort to generalize the self-consistent Ornstein-Zernike
  approximation (SCOZA) -- an accurate liquid-state theory that has
  been restricted so far to hard-core systems\nolinebreak -- to arbitrary
  soft-core systems we study a combination of SCOZA\ with a recently
  developed perturbation theory. The latter was constructed by
  Ben-Amotz and Stell [J. Phys. Chem. B 108,6877-6882 (2004)] as a
  reformulation of the Week-Chandler-Andersen perturbation theory
  directly in terms of an arbitrary hard-sphere reference system. We
  investigate the accuracy of the combined approach for the
  Lennard-Jones fluid by comparison with simulation data and pure
  perturbation theory predictions and determine the dependence of the
  thermodynamic properties and the phase behavior on the choice of the
  effective hard-core diameter of the reference system.\end{abstract}

\maketitle

\newpage

\section{Introduction}

\label{sec:intro}

In its most general formulation, the Self-Consistent Ornstein-Zernike
Approximation (SCOZA, \cite{scoza:12,scoza:11,scoza:9}) can be
obtained from a chosen liquid state theory by the introduction of an
adjustable parameter (such as an effective temperature) and the
subsequent imposition of consistency between two different routes to
thermodynamics, typically the energy and compressibility routes.
Simple as this prescription is, it has nevertheless proved to be
highly effective and to lead to very accurate predictions for
structure and thermodynamics throughout the temperature-density plane,
and for the critical point and phase coexistence in particular
\cite{scoza:1,scoza:8,scoza:6,plwk:bin:symm}.  Indeed, SCOZA\ is one
of only very few liquid state theories that do not develop serious
problems in the critical region and even exhibit some form of scaling
with non-classical, partly Ising-like critical exponents \cite{scoza:25}.

While this is by no means inherent to the concept of SCOZA,
applications to continuum fluids have been restricted in a number of
ways for both historical and practical reasons: Originally the theory
grew out of the semi-analytic solution of the Mean Spherical
Approximation (MSA) for hard core Yukawa potentials.
Correspondingly, SCOZA\ has so far only been used with an MSA-like
closure and with potentials composed of a hard core of diameter
$\sigma_\mathrm{H}$
and an attractive tail $ v_\mathrm{A}(r)$, a situation we will refer to as
the
``hard-attractive'' (HA) one.  As far as $ v_\mathrm{A}$ is concerned, the
original restriction to a single Yukawa term \cite{scoza:6} has gradually
been relaxed, first by expanding the class of admissible tails to
superpositions of two \cite{scoza:15} or more Yukawa terms
\cite{paschinger:dr},
\begin{displaymath}
v_\mathrm{A}(r) = \sum_\nu \epsilon_\nu\,\frac{\sigma_\mathrm{H}}
r\,e^{-z_\nu\,(r-\sigma_\mathrm{H})},
\end{displaymath}
and then to those of Sogami-Ise form \cite{scoza:28},
\begin{displaymath}
v_\mathrm{A}(r) =
\sum_\nu
\left(\epsilon_\nu^{(1)}\,\frac{\sigma_\mathrm{H}}r + 
\epsilon_\nu^{(2)}\right)
\, \,e^{-z_\nu\,(r-\sigma_\mathrm{H})},
\end{displaymath}
for which semi-analytic results are also available, only to be fully
overcome through the introduction of a fully numerical solution of the
Ornstein-Zernike (OZ) equation \cite{scoza:26}.  Approximation of
$ v_\mathrm{A}(r)$ as the superposition of a rather small number of Yukawa
or
Sogami-Ise tails --- a non-trivial step for some of the artificial
model potentials used in liquid state theory such as, \textit{e.~g.{}}, the
square
well one --- is thus no longer necessary, and it becomes possible to
use exactly the same interaction as the simulations one may want to
compare with.  Not surprisingly, the flexibility so gained comes at
substantial computational cost in solving the SCOZA\ partial
differential equation (PDE).

Despite this liberation of the form of the attractive tail,
application of SCOZA\ has so far remained tied to a hard core
reference fluid (marked by subscript H) with potential
\begin{equation} \label{eq:pot:h}
v_\mathrm{H}(r) =
\left\{
\begin{array}{ll}
\displaystyle +\infty&\displaystyle \hbox{for } r<\sigma_\mathrm{H}\\
\displaystyle 0&\displaystyle \hbox{for } r>\sigma_\mathrm{H}.
\end{array}
\right.
\end{equation}
Only in the case of a bounded interaction like, \textit{e.~g.{}}, the
Gaussian one,
$v(r) \propto \exp(-\alpha\,r^2)$, has the reference system
effectively been eliminated by letting $\sigma_\mathrm{H}$ go to zero
formally
\cite{scoza:27}.  But this is not an option for simple liquids where the
potential diverges for $r\to0$, such as the classic Lennard-Jones
(LJ) one,
\begin{equation} \label{eq:pot:lj}
v_\mathrm{LJ}(r) = 4\epsilon\,\left[\left(\frac\sigma r\right)^{12}
-\left(\frac\sigma r\right)^6\right].
\end{equation}
For this or more specific systems the infinitely strong hard core
repulsion $ v_\mathrm{H}(r)$ present in SCOZA\ is hardly realistic.  It is
therefore desirable to generalize the theory to largely arbitrary,
strong but soft short-range repulsion $ v_\mathrm{S}(r)$ to which a weak
attractive tail $ v_\mathrm{A}(r)$ is again added to obtain the full
``soft-attractive'' (SA) system,
\begin{equation} \label{eq:pot:sa}
v_\mathrm{SA}(r) \equiv v(r) =  v_\mathrm{S}(r) +  v_\mathrm{A}(r).
\end{equation}

Just as in an approximate SA-SCOZA\ recently proposed that relies on
the virial theorem to gauge the amplitude of the pair distribution
function (PDF) at the molecular surface \cite{ar:21}, in the present
contribution we also opt for the time-honored strategy of representing
the soft repulsive core by hard spheres of state-dependent diameter.
We therefore need explicit prescriptions for both the state-dependence
of this diameter and for any correction terms that may be needed to
account for the softness of the reference fluid.  In doing so there is
considerable latitude, and many different schemes have been proposed
and used in the past \cite{bas:analytical,bas:reformulation}.  For
example, the highly successful first order perturbation theory due to
Weeks, Chandler and Anderson (WCA, \cite{wca}) solves both of these
problems by determining $\sigma_\mathrm{H}$ as a function of the
temperature
$T=k_B/\beta$ and the particle density $\rho$ such that the Helmholtz
free energies of the soft core reference system and the hard spheres
coincide.  Determination of the diameter is then rather involved and
leads to a dependence not only on temperature but also on density.
Even though it has been argued that $\rho$ dependent $\sigma_\mathrm{H}$ is
fundamentally more appropriate \cite{tang}, the practical advantages of a
simple prescription that depends on temperature only often outweighs a
slight loss in accuracy.  Presumably this accounts for much of the
lasting popularity of the Barker-Henderson (BH) theory according to
which $\sigma_\mathrm{H}(\beta)$ is computed as \cite{barker:henderson}
\begin{equation} \label{eq:sig:bh}
\sigma_\mathrm{BH}(\beta) = \int_0^\infty\left(1-e^{-\beta\,
v_\mathrm{S}(r)}\right)\,dr.
\end{equation}

In this context a recent first order perturbation theory due to
Ben-Amotz and Stell (BAS, \cite{bas:reformulation,allg:27}) is of
particular interest: While WCA\ consider a general reference system
that is in turn mapped onto an effective hard sphere fluid to which
attractions are then added, BAS\ start from a pure hard sphere
reference fluid instead and add two terms to the free energy to
account for $ v_\mathrm{S}$ and $ v_\mathrm{A}$, respectively.  This hard
core WCA, or
BAS, theory has been shown to yield results of an accuracy comparable
to that of the original WCA\ theory while being insensitive to the
precise choice of $\sigma_\mathrm{H}$.  This makes it possible to use an
even
simpler temperature dependence of the effective diameter
$\sigma_\mathrm{H}$ than
that of Eq.~(\ref{eq:sig:bh}) without loss of accuracy
\cite{bas:reformulation}, and the resulting scheme combines the
advantages of the WCA\ and BH\ theories.

It is this BAS\ approach to the treatment of the reference system that
we here set out to combine with SCOZA\ along the lines of a suggestion
first put forward by Raineri and co-workers \cite{allg:27}.
After a short reminder of SCOZA\ and BAS\ theory and a presentation of
their combination (section~\ref{sec:theory}) we apply our BAS\ SCOZA\
to the LJ\ interaction~(\ref{eq:pot:lj}), comparing our results with
simulation data as well as the predictions of BAS\ theory
(section~\ref{sec:application}).  It turns out that we can well
reproduce earlier BAS\ results and, in fact, improve upon them in
slightly supercritical isotherms.  As far as phase coexistence is
concerned, however, the residual diameter dependence of the free
energy translates into an uncertainty of the pressure $P$ and chemical
potential $\mu$ that renders the coexistence curve more sensitive to
variations in $\sigma_\mathrm{H}$ than expected.  As shown in the
appendix, this is related to the divergence of the compressibility at
the critical point.  Without a criterion to fix $\sigma_\mathrm{H}(T)$
unambiguously, and in the face of further problems connected to the
spinodal of the underlying SCOZA\ computation if the hard-sphere
diameter of the reference system gets too small, the BAS\ SCOZA\
considered here is thus found to lack the faculty of predicting the
LJ\ phase diagram with the accuracy one has come to expect from SCOZA.
In our concluding remarks we consider the prospects for a SCOZA\
adapted to non-hard sphere reference systems more generally
(section~\ref{sec:conclusion}).

\section{BAS\ corrections to SCOZA}

\label{sec:theory}

The starting point of BAS\ theory, and hence also of our combined
BAS\ SCOZA, is the splitting~(\ref{eq:pot:sa}) of the full potential into
its
repulsive and attractive components.  In the present work we adopt the
conventional WCA\ prescription of separating them at the position
$ r_\mathrm{min}$ of the potential's minimum, \textit{i.~e.{}},
\begin{equation} \label{eq:pot:split}
\begin{array}{rl}
\displaystyle  v_\mathrm{S}(r)=&\displaystyle \left\{
\begin{array}{ll}
\displaystyle v(r) - v( r_\mathrm{min})&\displaystyle \displaystyle
\hbox{for } r <  r_\mathrm{min}\\
\displaystyle 0&\displaystyle \displaystyle \hbox{otherwise}
\end{array}\right.\\
\displaystyle  v_\mathrm{A}(r)=&\displaystyle \left\{
\begin{array}{ll}
\displaystyle v( r_\mathrm{min})&\displaystyle \displaystyle \hbox{for } r
<  r_\mathrm{min}\\
\displaystyle v(r)&\displaystyle \displaystyle \hbox{otherwise;}
\end{array}\right.
\end{array}
\end{equation}
in the LJ\ case, $ r_\mathrm{min} = \sqrt[6]{2}\,\sigma$ and $v(
r_\mathrm{min}) =
-\epsilon$.  The Helmholtz free energy of the fluid is customarily written
as the sum of the ideal gas contribution $ A^\mathrm{id}$ and the excess
free
energy $ A^\mathrm{ex}$,
\begin{displaymath}
A =  A^\mathrm{id} +  A^\mathrm{ex}.
\end{displaymath}
In computing $ A^\mathrm{ex}$, BAS\ start from a hard sphere reference
system
and split $ A^\mathrm{ex}$ into the hard sphere term $
A^\mathrm{ex}_\mathrm{H}$ and two correction
terms corresponding to $ v_\mathrm{S}$ and $ v_\mathrm{A}$, respectively,
\begin{displaymath}
A^\mathrm{ex} =  A^\mathrm{ex}_\mathrm{H} + \Delta A_\mathrm{S} + \Delta
A_\mathrm{A}.
\end{displaymath}

The term $\Delta A_\mathrm{S}$ implementing the difference between hard (H)
and soft
(S) repulsion is given by \cite{bas:reformulation}
\begin{equation} \label{eq:dAs}
\frac{\beta\Delta A_\mathrm{S}}N
= 2\pi\rho\int_0^\infty\left( g_\mathrm{H}(r)-
g_\mathrm{S}(r)\right)\,r^2\,dr,
\end{equation}
where the $g(r)$ denote the respective PDFs.  As in WCA\ and BAS\
theories, we relate $ g_\mathrm{S}$ to $ g_\mathrm{H}$ by the assumption of
equal cavity
function $y(r) = g(r) \* \exp\left(+\beta \* v(r)\right)$.  This is
expected to be quite harmless for the rather stiff LJ\ case
considered here \cite{bas:hard}.  We have checked that the three-term
approximation for $y_\mathrm{H}(r)$ inside the core used in
Ref.~\cite{bas:analytical} gives essentially the same results as a
four-term one due to Grundke and Henderson \cite{hs:2}.  Outside the core
we adopt Waisman's prescription for the hard sphere structure
functions: The direct correlation function $ c_\mathrm{H}(r)$,
$r>\sigma_\mathrm{H}$, is
approximated by a single Yukawa tail with parameters chosen so as to
reproduce the Carnahan Starling equation of state through both the
compressibility and the virial routes \cite{waisman}, \textit{cf.{}}\
appendix~A of
Ref.~\cite{scoza:6}.

The remaining contributions to the free energy are $
A^\mathrm{ex}_\mathrm{H}$ and $\Delta A_\mathrm{A}$.
For the former one may use the well-known Carnahan-Starling result
\cite{hansen:mcdonald}, whereas the latter is given in BAS\ theory as an
integral over $ g_\mathrm{H}(r)\, v_\mathrm{A}(r)$
\cite{bas:reformulation}.  On the other hand, $ A^\mathrm{ex}_\mathrm{HA} =
A^\mathrm{ex}_\mathrm{H} + \Delta A_\mathrm{A}$ corresponds to the
superposition of hard core repulsion
and long-range attraction, \textit{i.~e.{}}, to the potential
\begin{displaymath}
v_\mathrm{HA}(r) =  v_\mathrm{H}(r) +  v_\mathrm{A}(r).
\end{displaymath}
This is precisely the HA\ situation SCOZA\ has long been applied to
with excellent results.  In combining BAS\ theory and SCOZA\ it is
therefore natural to use SCOZA\ for the description of this HA\
sub-problem, and for the computation of $ A^\mathrm{ex}_\mathrm{HA}$ in
particular.  By
adding $\Delta A_\mathrm{S}$ we then arrive at the full excess free energy,
from
which any remaining thermodynamic quantities may be derived.  This
recipe was first proposed in Ref.~\cite{allg:27}.

The first step in computing the thermodynamics of the full SA\ system
along an isotherm at inverse temperature $\beta$ within the combined
BAS\ SCOZA\ is the determination of the effective hard sphere diameter
$\sigma_\mathrm{H}$.  In view of the purported insensitivity of BAS\ theory
results to the precise choice of $\sigma_\mathrm{H}$, we generally employ a
simple
Boltzmann factor criterion (BFC) to fix the temperature dependence of
$\sigma_\mathrm{H}$ \cite{bas:reformulation,allg:27}.  According to this
criterion,
we choose $\sigma_\mathrm{H}(\beta)$ so as to keep $\beta\*
v_\mathrm{S}(\sigma_\mathrm{H})$ constant,
$\exp(-\beta\* v_\mathrm{S}(\sigma_\mathrm{H})) = 1/ a_\mathrm{BFC}$.  Here
we have introduced a
parameter $ a_\mathrm{BFC}$ whose variation allows us to change
$\sigma_\mathrm{H}$ in a
consistent way for all temperatures.  In Ref.~\cite{bas:reformulation},
BAS\ found virtually unchanged results for the parameter range
$2\le a_\mathrm{BFC}\le5$, which includes the \textit{a priori} preferred
value
$ a_\mathrm{BFC}=e$ \cite{allg:27}. --- In some of the computations we also
use the
BH\ prescription~(\ref{eq:sig:bh}).

With $\sigma_\mathrm{H}$ so determined, the SCOZA\ part can be handled in
the
usual way \cite{paschinger:dr} except that the WCA\
splitting~(\ref{eq:pot:split}) effectively precludes a multi-Yukawa or
Sogami-Ise representation of $ v_\mathrm{A}(r)$ for $r>\sigma_\mathrm{H}$. 
We thus have no
choice but to turn to an implementation where the OZ\ relation, along
with the customary SCOZA\ closure
\begin{equation} \label{eq:scoza:closure}
\begin{array}{rll}
\displaystyle  g_\mathrm{HA}(r)&\displaystyle =0&\displaystyle \hbox{for
}r<\sigma_\mathrm{H}\\
\displaystyle  c_\mathrm{HA}(r)&\displaystyle = c_\mathrm{H}(r) - K\,
v_\mathrm{A}(r)&\displaystyle \hbox{for }r>\sigma_\mathrm{H},
\end{array}
\end{equation}
is solved numerically \cite{scoza:26}.  The first of the above relations
expresses the impenetrability of the effective hard cores whereas the
second one describes the effect of the attractive tail on the HA\
direct correlation function $ c_\mathrm{HA}(r)$ outside the core beyond
that of
hard spheres, $ c_\mathrm{H}(r)$.  Comparison immediately shows that this
corresponds to the MSA\ closure with $\beta$ replaced by a state
dependent effective inverse temperature $K(\beta,\rho)$.  The latter
is fixed through the requirement of thermodynamic self-consistency as
embodied in the SCOZA\ PDE
\begin{equation} \label{eq:scoza:pde}
\left(\frac\partial{\partial\beta}\frac1{\chi_\mathrm{HA}}\right)_\rho
= \rho\,\left(\frac{\partial^2 u_\mathrm{HA}}{\partial\rho^2}\right)_\beta.
\end{equation}
Here $1/\chi_\mathrm{HA} = 1 - \rho \* \int_{\mathbb{R}^3}
c_\mathrm{HA}(r)\*d^3r$ so that
$\kappa = \beta \chi_\mathrm{HA} / \rho$ is the isothermal compressibility
of
the HA\ system as evaluated from the compressibility route, and $
u_\mathrm{HA}
\equiv  U^\mathrm{ex}_\mathrm{HA}/V = 2\pi \* \rho^2 \*
\int_{\sigma_\mathrm{H}}^\infty  g_\mathrm{HA}(r) \*
v_\mathrm{HA}(r) \* r^2dr$ is the energy route result for the excess, or
configurational, internal energy per unit volume.  Solution of this
PDE\ subject to suitable boundary conditions at infinite temperature
as well as at vanishing and high density yields both the structural
and the thermodynamic properties of the HA\ system throughout the
domain of the PDE.  Only below the critical temperature is there a
region of instability or non-convergence that must be excluded from
the integration.  It is customary to do so through the imposition of
an additional boundary at the spinodal so that no SCOZA\ results are
available inside the HA\ spinodal.  For a detailed description of the
numerical procedure we refer the reader to the literature
\cite{scoza:8,paschinger:dr,scoza:26}.

For use in BAS\ SCOZA, integration of the SCOZA\ PDE\ is stopped upon
reaching the temperature corresponding to $\sigma_\mathrm{H}$, and internal
energy,
free energy, pressure, and chemical potential of the HA\ system are
extracted along this isotherm.  Finally, we add the BAS\ correction
terms following from $\Delta A_\mathrm{S}$ in order to arrive at the final
SA\
results for these quantities.

Determination of the SA\ phase diagram involves repeating this
procedure for all the diameters corresponding to the temperature range
of interest.  As both the HA\ free energy so obtained and the BAS\
correction term $\Delta A_\mathrm{S}$ are unique functions of temperature
and
density, so is their sum $ A^\mathrm{ex}$, the excess free energy of the
full
SA\ system.  The SCOZA\ PDE\ is then trivially fullfilled as long
as the thermodynamic quantities are computed by differentiation of
$ A^\mathrm{ex}$.  On the other hand, the structure of the SA\ system
differs from that of
the HA\ system computed with SCOZA\ and is therefore not accessible.
Consequently, neither the compressibility route nor the energy route
to thermodynamics can be evaluated for the SA\ case within the
present approach.  In this sense, BAS\ SCOZA\ must not be seen as a soft
core version of SCOZA\ but rather as a BAS\ theory built upon a
SCOZA\ foundation.

\section{Application to the LJ\ fluid}

\label{sec:application}

In order to gauge the performance of BAS\ SCOZA\ as described in the
preceding section we now turn to its application to the LJ\
potential~(\ref{eq:pot:lj}) with the WCA\ splitting~(\ref{eq:pot:split}). 
For
comparison purposes we make use of the molecular dynamics simulation
data published by Johnson and co-workers \cite{gubbins} and by Lotfi and
co-workers \cite{lj:3} as well as Monte Carlo (MC) results due to
Wilding \cite{wilding}.  While the former two directly relate to the full
interaction $ v_\mathrm{LJ}$ of Eq.~(\ref{eq:pot:lj}), the latter are for
the cut but
not shifted potential $ v_\mathrm{LJ}(r) \* \Theta(2.5\sigma-r)$ without
corrections ($\Theta$ is Heaviside's function).  We have therefore
exploited the flexibility brought about by the fully numerical
solution of the OZ\ relation and performed BAS\ SCOZA\ computations with
both of these potentials, depending on the simulation data we want to
compare with.

In presenting their theory \cite{bas:reformulation,allg:27} as well as
in their comparative study of various thermodynamic perturbation
theories \cite{bas:analytical}, BAS\ focussed on high densities, and
mostly on rather high temperatures: With only a single subcritical
temperature and with densities that are at least twice the critical
one, all of the four states repeatedly considered in
Ref.~\cite{bas:reformulation} are quite far from the critical point and
the spinodal.  Presumably they were chosen because it is in dense or
hot systems that particles explore distances around $\sigma_\mathrm{H}$
effectively
so that high temperatures and densities present the most challenging
test for a perturbative description of soft cores.  The effects of
variations of $\sigma_\mathrm{H}$ generally increase with density for the
isotherms
studied in Ref.~\cite{bas:reformulation} as well as in our own
calculations.

For a comparison of BAS\ SCOZA\ with pure BAS\ theory (or actually,
$\Delta A_\mathrm{S}$ being identical in both cases, of their HA\ parts
only) these
states are only of limited interest: At high density, the pair
structure is dominated by packing effects so that any difference
between $\beta$ and the parameter $K$ of the
closure~(\ref{eq:scoza:closure}) hardly affects the HA\ energy integral
and the free energy obtained from it by thermodynamic integration.
For the lower density states, on the other hand, the high temperatures and
great
separations from the critical region imply that the SCOZA\
self-consistency problem can hardly have changed $K/\beta$ from unity
appreciably and so again render the HA\ structure and thermodynamics
of BAS\ SCOZA\ equivalent to pure BAS\ theory.  Any remaining
discrepancies in the thermodynamics must be attributed to the
difference between first-order perturbation theory and what is
essentially the MSA\ for the HA\ problem.  It is therefore no
surprise that BAS\ SCOZA\ reproduces the BAS\ results for these states
very well.  A detailed comparison of the data underlying Fig.~\ref{fig:Aex}
with the corresponding Fig.~2 of Ref.~\cite{bas:reformulation} does,
however, seem to hint at a sensitivity of $ A^\mathrm{ex}$ to variation of
$\sigma_\mathrm{H}$ that is slightly reduced \textit{vis-\`a-vis}\ BAS\
theory at
$ T^* 
\equiv k_B T/\epsilon = 2.81$, $ \rho^* 
\equiv \rho\,\sigma^3 = 0.85$, and slightly increased for the smallest
$\sigma_\mathrm{H}$ values considered at $ T^* = 0.75$, $ \rho^* = 0.84$;
for the
latter temperature, however, $\sigma_\mathrm{H}\sim0.94$ corresponds to $
a_\mathrm{BFC} \sim
10^2$, which is far outside the normal range, $2 \le  a_\mathrm{BFC}
\le 5$.  For the two remaining states, $ T^* = 3.05$, $ \rho^* = 1.1$ and
$ T^* = 1.35$, $ \rho^* = 0.65$, any differences are too small to be made
out from Fig.~2 of Ref.~\cite{bas:reformulation}.

SCOZA\ differs from MSA\ mainly in the critical region and in the
vicinity of the spinodal where $K$ strongly deviates from $\beta$.  In
order to see a genuine SCOZA\ contribution, as opposed to merely
gauging the accuracy of a first order perturbation theory for the HA\
subproblem, our interest is naturally drawn closer to the HA\
critical point, and thus also closer to the critical point of the full
SA\ system.  Of the isotherms displayed in Fig.~4 of
Ref.~\cite{bas:reformulation}, the one at $ T^* = 1.35$ should already be
sufficiently close to the critical temperature of the full LJ\
interaction that has been estimated as $  T_c^* = 1.310$ \cite{lj:3} and
1.313 \cite{gubbins} by molecular dynamics, and $  T_c^* = 1.3120(7)$
\cite{lj:2} and 1.326(2) \cite{caillol} by Monte Carlo methods.
Unfortunately, BAS\ only show $Z=\beta P/\rho$ in the figure so that
it is difficult to discern whether a van der Waals loop is present,
nor do they address this question directly.  On the other hand, the
standard by
which the performance of BAS\ theory is judged in
Ref.~\cite{bas:reformulation} is the classic WCA\ theory.  As all of BH\
theory, WCA\ theory, and a thermodynamically self-consistent variant
of the latter due to Lado display almost identical van der Waals loops
at even higher temperature ($ T^* = 1.4$, \textit{cf.{}}\ inset in Fig.~5
of
Ref.~\cite{bas:analytical}), we may safely assume that this is true for
the BAS\ result, too.  As can be seen from the inset in our
corresponding Fig.~\ref{fig:P}, the same isotherm is correctly predicted to
lie above the critical temperature in BAS\ SCOZA\ and is in essentially
perfect aggreement with the simulation data for $2.3 \le  a_\mathrm{BFC}
\le e$.

By way of contrast, the same diameters give substantial deviations and
show a marked trend towards a van der Waals loop for smaller $
a_\mathrm{BFC}$
when BAS\ SCOZA\ is replaced by ``BAS\ MSA'', \textit{i.~e.{}}, if $K$ is
restricted
to coincide with $\beta$, \textit{cf.{}}\ inset in Fig.~\ref{fig:P}.  While
an
adjustment of $ a_\mathrm{BFC}$ so as to achieve good agreement with the
simulation data in this part of the phase diagram is certainly
possible, the resulting diameter may be too small to describe the
packing effects at high density correctly.  At any rate, for slightly
supercritical isotherms like that at $ T^* = 1.4$, BAS\ SCOZA\ not only
leads to a better agreement with the pressure data but also proves
less sensitive to variations of the diameter.  It is therefore
superior not only to pure BAS\ theory (which might be explained by
the qualitative difference of perturbation theory \textit{vs.{}}\ integral
equations) but also to MSA\ with BAS\ corrections.  The improvements
must therefore be attributed to SCOZA's self-consistency requirement.

Returning to BAS\ SCOZA, even for this narrow $ a_\mathrm{BFC}$ range there
appears
a systematic, if small, pressure difference that increases with
density.  Although it is not obvious how to relate the sensitivities
to variation of $\sigma_\mathrm{H}$ of different quantities, comparison of
Figs.~\ref{fig:P} and~\ref{fig:A:ex} seems to indicate that the pressure
depends
on $\sigma_\mathrm{H}$ more strongly than the free energy.  A similar
conclusion
can be drawn for the chemical potential by comparing Figs.~\ref{fig:Aex}
and~\ref{fig:mu}.  This greater sensitivity of $P$ and $\mu$ relative to
$ A^\mathrm{ex}$ for most states, and for comparatively low temperatures in
particular, is also demonstrated in Tab.~\ref{tab:sensitivity} where
$\sigma_\mathrm{H}$
induced changes are related both to the absolute value and to the
density dependence of the respective quantities.  It can also be seen
in pure BAS\ theory: For the lowest isotherm in Fig.~4 of
Ref.~\cite{bas:reformulation}, $ T^* = 0.74$, a slight variation of
$\sigma_\mathrm{H}$
from $\sigma_\mathrm{BFC}^{[a=e]}$ by only two per cent is sufficient even
to change
$Z$ from a convex to a concave function of density for
$ \rho^* \approx 1$.

Proceeding to the critical region and phase coexistence, in BAS\ SCOZA\
the binodal is located as in pure SCOZA\, \textit{viz.{}}, by a search for
densities of equal pressure $P$ and chemical potential $\mu$; the
critical point is identified with the locus where the gas and liquid
branches of the binodal meet.  As expected for a variant of BAS\
theory even if it makes use of SCOZA\ input, the critical behavior is
not compatible with the Ising universality class: In the accessible
temperature range the coexistence curve is not described well by the
usual scaling form, and for the highest sub-critical temperatures the
effective exponent $\beta$ tends to values far larger than SCOZA's
$7/20$ \cite{scoza:25}, which in turn slightly exceeds the correct Ising
value.  This is also conspicuous from the forms of the binodals
obtained from the BFC\ diameter $\sigma_\mathrm{BFC}$ with several values
of
$ a_\mathrm{BFC}$ as well as the BH\ diameter $\sigma_\mathrm{BH}$ as
displayed in
Figs.~\ref{fig:coex:tr} and~\ref{fig:coex:or} for the truncated and the
full
LJ\ potentials, respectively.

Figs.~\ref{fig:coex:tr} and~\ref{fig:coex:or} also bring out the gravity of
the diameter sensitivity of $P$ and $\mu$ for the description of phase
coexistence: There is a substantial variation in the locations of the
upper parts of the coexistence curves when $\sigma_\mathrm{H}$ is changed
from
$\sigma_\mathrm{BFC}^{[a=2.2]}$ (Fig.~\ref{fig:coex:tr}) or
$\sigma_\mathrm{BFC}^{[a=2.3]}$
(Fig.~\ref{fig:coex:or}).  It should be noted that these optimal values of
$ a_\mathrm{BFC}$ cannot be computed from the theory itself but are merely
the
results of comparisons with the simulation data for the coexisting
densities for the truncated and full potentials, respectively.

The particularly great sensitivity of the binodal at the highest
temperatures can in fact easily be understood by linking the shift in
the coexisting densities induced by a change in the effective hard
core diameter $\sigma_\mathrm{H}$ to the isothermal compressibility
$\kappa$ at
phase coexistence.  Clearly, $\kappa$ diverges at the critical point
and then decreases along the two branches of the binodal as we proceed
to lower temperatures.  Informally speaking, a higher compressibility
means that the physical system must be compressed or expanded, and
therefore its density changed, by a larger amount in order to offset a
small change in pressure and chemical potential, which explains the
more pronounced effect at higher temperatures.  At the same time, the
thermodynamic quantities are generally more sensitive to a change in
$\sigma_\mathrm{H}$ at higher densities, which explains the asymmetry of
the effect
between the low and high density branches of the binodal.  A more
formal exposition of this reasoning can be found in the appendix.

A different perspective on the strong $\sigma_\mathrm{H}$ dependence of the
binodal
is offered by the realization that an increase in $\sigma_\mathrm{H}$
renders the
HA\ subproblem more strongly repulsive, thus lowering the HA\
critical temperature.  In an exact theory, the correction term $\Delta
A_\mathrm{S}$
strictly compensates this shift of the critical point.  From a
first-order perturbation theory, however, we can expect only a partial
compensation, as is clearly demonstrated in Fig.~\ref{fig:deltaA} where we
separate the SCOZA\ and BAS\ contributions to the free energy.
Consequently, a change in $ a_\mathrm{BFC}$ (or, more generally, in the
prescription for $\sigma_\mathrm{H}(\beta)$) still leads to a systematic
shift of
the SA\ critical temperature.  For any given isotherm, this
necessarily corresponds to changes of the coexisting densities that
are most pronounced close to the critical point where the binodal
gains in width most rapidly.

Such shifts of $ \beta_c$ can actually be inferred from
Figs.~\ref{fig:coex:tr} and~\ref{fig:coex:or}.  The reason they cannot be
seen
directly is connected to the specific way in which BAS\ SCOZA\ combines
the theories it is built upon, \textit{viz.{}}, by adding a BAS\ correction
term
to a SCOZA\ free energy: As mentioned in the introduction, SCOZA\
suffers from a region where no solution can be found due to either
stability or convergence problems.  This region is buried within the
HA\ spinodal and is customarily eliminated through the imposition of
an artificial spinodal boundary condition.  No BAS\ SCOZA\ results can
therefore be obtained inside the HA\ spinodal.  Depending on the
diameter, this hole in the solution may lie well within the SA\
binodal.  For $ a_\mathrm{BFC}$ above a certain threshold, however, or more
generally whenever $\sigma_\mathrm{H}( \beta_c)$ is too small, the HA\
interaction
is not sufficiently repulsive so that the HA\ spinodal rises above
the SA\ binodal, the upper part of which is then no longer
accessible.  As can be seen from Figs.~\ref{fig:coex:tr}
and~\ref{fig:coex:or},
this is the case for those diameters that come close to the simulated
phase diagrams.  The hole in the solution for temperatures below the
HA\ critical one can also be seen from the missing intervals in the
lowest temperature isotherms displayed in Figs.~\ref{fig:P}, \ref{fig:A:ex}
and~\ref{fig:U}.

Let us now consider the configurational internal energy $ U^\mathrm{ex} =
\left(\partial\beta A^\mathrm{ex} / \partial\beta \right)_\rho$.  With $
A^\mathrm{ex} =
A^\mathrm{ex}_\mathrm{HA} + \Delta A_\mathrm{S}$ we obtain $ U^\mathrm{ex}$
as the sum of the HA\ internal energy
$ U^\mathrm{ex}_\mathrm{HA}$ extracted directly from the SCOZA\ computation
with fixed
diameter and the correction $\Delta U_\mathrm{S}$ implementing the
difference between
hard and soft repulsion.  The latter is composed of the temperature
derivative of the BAS\ correction $\Delta A_\mathrm{S}$ and a term related
to the
temperature dependence of $ A^\mathrm{ex}_\mathrm{HA}$ through
$\sigma_\mathrm{H}$ (and thus to the
steepness of $ v_\mathrm{S}$ at the effective molecular surface),
\begin{displaymath}
\begin{array}{rl}
\displaystyle  U^\mathrm{ex} &\displaystyle =  U^\mathrm{ex}_\mathrm{HA} +
\Delta U_\mathrm{S}\\
\displaystyle \Delta U_\mathrm{S} &\displaystyle =
\left(\frac{\partial\beta\Delta A_\mathrm{S}}{\partial\beta}\right)_\rho
+ \left(\frac{\partial\beta
A^\mathrm{ex}_\mathrm{HA}}{\partial\sigma_\mathrm{H}}\right)_\beta
\, \frac{{{\mathrm d}}\sigma_\mathrm{H}}{{{\mathrm d}}\beta}.
\end{array}
\end{displaymath}
Here, the derivative ${{{\mathrm d}}\sigma_\mathrm{H}}/{{{\mathrm
d}}\beta}$ is to be evaluated according
to the convention used to fix $\sigma_\mathrm{H}(\beta)$, \textit{i.~e.{}},
from
Eq.~(\ref{eq:sig:bh}) for $\sigma_\mathrm{BH}$, or at constant $
a_\mathrm{BFC}$ for $\sigma_\mathrm{BFC}$.

From Fig.~\ref{fig:Uex} we see that the overall diameter dependence of the
excess internal energy is qualitatively similar to that of the excess
free energy, \textit{cf.{}}\ Fig.~\ref{fig:Aex}.  Not surprisingly, it is
most
pronounced at the highest densities whereas the temperature is of
minor importance for the states considered in Fig.~\ref{fig:Uex}.  This can
easily be understood in terms of packing effects that dominate the
pair structure.  In Fig.~\ref{fig:U:split} we again separate the SCOZA\
and BAS\ contributions to the internal energy.  Interestingly, the
latter compensates the diameter dependence of the former not even
approximately as was the case for the free energy (Fig.~\ref{fig:deltaA}).
Instead, $\Delta U_\mathrm{S}$ is essentially constant over the whole
$\sigma_\mathrm{H}$ range
shown, and the variation of $ U^\mathrm{ex}$ is due to $
U^\mathrm{ex}_\mathrm{HA}$ alone.  As
Fig.~\ref{fig:U} shows, the particularly low sensitivity of $
U^\mathrm{ex}$ with
respect to $\sigma_\mathrm{H}$ in the upper panel of Fig.~\ref{fig:U:split}
is a mere
consequence of the transition between density ranges where $ U^\mathrm{ex}$
rises or falls, respectively, with growing $\sigma_\mathrm{H}$.  In the
lower
panel, the internal energy again varies by several per cent in the
$\sigma_\mathrm{H}$ range displayed.

Both pressure $P$ and internal energy $ U^\mathrm{ex}$ as computed within
BAS\ SCOZA\ with $2.3 \le  a_\mathrm{BFC} \le e$ are generally in good
agreement
with the simulation results for the sample isotherms displayed in
Figs.~\ref{fig:P} and~\ref{fig:U}.  Only a detailed comparison reveals some
systematic deviations that seem to hint at a moderate state dependence
of the optimal value of $ a_\mathrm{BFC}$: At high densities and
supercritical
temperatures $ U^\mathrm{ex}$ is best reproduced with $ a_\mathrm{BFC}=e$,
and the
simulated pressures indicate an $ a_\mathrm{BFC}$ even slightly larger than
$e$.
By way of contrast, phase diagram (Fig.~\ref{fig:coex:or}), pressure
(Fig.~\ref{fig:P}) and internal energy (Fig.~\ref{fig:U}) all agree that
$ a_\mathrm{BFC}=2.3$ is the optimum value for the full LJ\ interaction for
states for which SCOZA's self-consistency requirement is expected to
be relevant, \textit{i.~e.{}}, for $T <  T_c$ as well as for $T \sim  T_c$,
$\rho
\sim  \rho_c$.  A similar level of agreement is also expected for the
truncated potential with a slightly larger diameter, $ a_\mathrm{BFC}=2.2$,
\textit{cf.{}}\
Fig.~\ref{fig:coex:tr}.  This difference of about 0.1~per cent in
$\sigma_\mathrm{H}$
seems remarkably small given the pronounced influence on the critical
parameters such a cut may have \cite{smit}, and in particular given the
respective critical temperatures $  T_c^* < 1.2$ for the cut
potential \cite{wilding} as opposed to $  T_c^* > 1.3$ for the full
interaction \cite{gubbins,lj:3,lj:2,caillol}, \textit{cf.{}}\
Figs.~\ref{fig:coex:tr}
and~\ref{fig:coex:or}.

\section{Conclusion and perspectives}

\label{sec:conclusion}

As we argued at the end of section~\ref{sec:theory}, BAS\ SCOZA\ should be
regarded as a SCOZA-based variant of BAS\ theory rather than as a
modified SCOZA.  From this point of view our expectations regarding
the performance of this combined theory are well fulfilled: Good
agreement with simulation results for binodal, pressure and internal
energy can be achieved with a judicious choice of the parameter
$ a_\mathrm{BFC}$; we find a significant improvement over pure BAS\ theory
at
slightly supercritical isotherms; and away from the critical region
and phase coexistence both schemes give essentially equivalent
results.  Unfortunately, the transition from pure BAS\ theory to
BAS\ SCOZA\ entails a dramatic increase in computational cost as the
solution of a non-linear PDE\ in $\beta$ and $\rho$ is required just
for the results along a single isotherm.  For the computation of a
full phase diagram, BAS\ SCOZA\ theory is far more demanding than even
fixed-diameter SCOZA\ where a single integration of the PDE\ yields
the results at all densities and temperatures.

On the other hand, the present scheme was originally proposed as a
soft-core extension of SCOZA\ rather than as an improved BAS\ theory
\cite{allg:27}.  And indeed, BAS\ SCOZA\ is able to handle SA\ fluids
using the true interaction whereas previous applications of
conventional SCOZA\ to the LJ\ fluid had to rely on heuristic, and by
no means small, corrections to form and amplitude of the potential
outside the fixed hard core \cite{scoza:15} yielding a critical
temperature that is too low by about five per cent
\cite{paschinger:dr}.  Still, the huge increase in computational cost,
the problems connected to the HA\ spinodal, and the loss of all
structural information may seem an inordinate prize to pay for this
advantage of BAS\ SCOZA.

Most importantly, though, the distinctive feature of SCOZA,
\textit{viz.{}}, the
thermodynamic consistency requirement implemented by an effective
temperature that replaces the true temperature in the MSA\ closure
for the HA\ system, shows up in the results only in the proximity of
the critical point.  It is there that we have to look for
SCOZA-specific improvements: In the rest of the $(\beta,\rho)$ plane
SCOZA\ is essentially equivalent to MSA, and any improvements over
pure BAS\ theory merely reflect the superiority of integral equations
over perturbation theory in evaluating the difference $\Delta A_\mathrm{A}$
of the
free energies of the H\ and HA\ systems.  Close to the critical
point, however, the chief advantage of BAS\ theory, \textit{viz.{}}, the
near-constancy of the thermodynamic results under variation of
$\sigma_\mathrm{H}$
breaks down, and the BAS\ SCOZA\ binodal strongly depends on
$\sigma_\mathrm{H}(\beta)$.  The value of the parameter $ a_\mathrm{BFC}$
thus greatly
influences the predicted phase behavior; at the same time, it cannot
be computed from the theory itself, and the \textit{a priori}
attractive choice of $ a_\mathrm{BFC} = e$ \cite{allg:27} is certainly
unsatisfactory in the critical region.  Further considering that the
BAS\ SCOZA\ binodals are far from compatible with the Ising universality
class we therefore conclude that the theory as presented here cannot
be used to predict the critical properties and liquid-vapor phase
behavior from first principles with anywhere near the accuracy usually
associated with SCOZA.

As for the feasability of a BAS-based SCOZA\ rather than a
SCOZA-based BAS\ theory, the main obstacle to using BAS\ results to
``provide reference-system input'' \cite{bas:analytical} for SCOZA\ is
the purely thermodynamic nature of the BAS\ term $\Delta A_\mathrm{S}$ that
does not
allow us to relate the structures of the HA\ and SA\ fluids.
Without $K$-dependent SA\ structures, however, neither the energy nor
the compressibility routes to the SA\ thermodynamics can be
evaluated, and the question of their consistency becomes meaningless.
Indeed, inclusion of the soft core contribution in the thermodynamic
self-consistency problem is likely to be of prime importance for the
performance of an eventual soft core variant of SCOZA\ \cite{ar:21}.

What then, one may ask, are more promising ways towards a SCOZA\
capable of describing SA\ systems?  For this we see two options: One
possibility is to consider formulations where the introduction of an
effective hard core reference system still allows the analytical short
cuts for the solution of the OZ\ equation available for the
multi-Yukawa or Sogami-Ise fluids to be used and that include the
softness in the self-consistency problem, if only approximately; one
such approach has recently been proposed \cite{ar:21}.  On the other hand,
if the
hard-won freedom to choose almost arbitrary functions $ v_\mathrm{A}$ and $
v_\mathrm{S}$
for the attractive and soft-repulsive parts of the interaction is to
be retained, the OZ\ relation must be solved numerically at any rate.
In this case there is no need for a hard core reference system, and
non-perturbative ways of treating the soft repulsive reference fluid
may be more appropriate and less wasteful in terms of computing
resources and structural information.  We intend to pursue this topic
in a further study.

\section*{Acknowledgments}

AR\ wishes to thank Johan H\o ye for many stimulating discussions and
gratefully acknowledges financial support from the Austrian Science
Fund (FWF) under project~J2380-N08.  This work was also supported by
the Austrian Science Fund (FWF) under Project No.~P17178-N02.

\appendix
\renewcommand{\theequation}{A\arabic{equation}}\setcounter{equation}{0}

\section{Diameter sensitivity of the binodal}

Let us consider a subcritical isotherm, $\beta> \beta_c$, with vapor and
liquid coexisting at densities $ \rho_v$ and $ \rho_l$, respectively.
Along this isotherm the exact free energy is a function of density
only, $A = A(\rho)$.  The BAS\ free energy, on the other hand, is computed
in an
approximate way and makes use of an effective hard sphere reference
system with density independent diameter $\sigma_\mathrm{H}$.  It therefore
acquires an additional dependence.  To obtain the free energy at
slightly different values of $\sigma_\mathrm{H}$ and $\rho$ we can use the
first
order Taylor expansion
\begin{displaymath}
A(\rho+\Delta\rho,\sigma_\mathrm{H}+\Delta\sigma_\mathrm{H}) = 
A(\rho) + \Delta\sigma_\mathrm{H}\,\dot A(\rho) + \Delta\rho\,A'(\rho) +
\ldots\,.
\end{displaymath}
In this appendix we use dots to denote differentiation with respect to
$\sigma_\mathrm{H}$ and primes for differentiation with respect to $\rho$. 
For the
benefit of a compact notation we will also use subscripts $v$ and $l$
to indicate functions evaluated at $ \rho_v$ and $ \rho_l$, respectively,
as well as subscripts $+$ or $-$ for their symmetric and
anti-symmetric combinations, \textit{i.~e.{}}, $\psi_\pm \equiv \psi_l \pm
\psi_v
\equiv \psi( \rho_l) \pm \psi( \rho_v)$ for any quantity $\psi$.

By differentiation of the free energy we get analogous relations for
the pressure $P$ and chemical potential $\mu$, the two quantities that
directly enter the determination of the coexistence curve:
\begin{displaymath}
P(\rho+\Delta\rho,\sigma_\mathrm{H}+\Delta\sigma_\mathrm{H}) =
P(\rho) + \Delta\sigma_\mathrm{H}\,\dot P(\rho) + \Delta\rho\, P'(\rho) +
\ldots,
\end{displaymath}
\begin{displaymath}
\mu(\rho+\Delta\rho,\sigma_\mathrm{H}+\Delta\sigma_\mathrm{H}) =
\mu(\rho) + \Delta\sigma_\mathrm{H}\,\dot\mu(\rho) +
\Delta\sigma_\mathrm{H}\,\mu'(\rho) + \ldots\,.
\end{displaymath}

For given $\sigma_\mathrm{H}$, the coexisting densities are obtained from
the
criterion of equal pressure and chemical potential, $P(
\rho_v,\sigma_\mathrm{H}) =
P( \rho_l,\sigma_\mathrm{H})$ and $\mu( \rho_v,\sigma_\mathrm{H}) = \mu(
\rho_l,\sigma_\mathrm{H})$.  Specializing
to the coexisting densities $ \rho_v$ and $ \rho_l$ obtained for this
$\sigma_\mathrm{H}$, the equilibrium conditions reduce to $P_v = P_l$ and
$\mu_v =
\mu_l$, or
\begin{equation} \label{eq:ref:equil:p:mu}
P_- = \mu_- = 0.
\end{equation}
Non-zero $\Delta\sigma_\mathrm{H}$, on the other hand, modifies both
pressure and chemical potential
and is therefore generally accompanied by changes $\Delta\rho_v$ and
$\Delta\rho_l$ in the coexisting densities.  Taking
Eq.~(\ref{eq:ref:equil:p:mu}) into account, switching to symmetric and
anti-symmetric combinations of quantities, and simplifying, the
equilibrium conditions are easily obtained as
\begin{displaymath}
P'_-\,\Delta\rho_+ +  P'_+\,\Delta\rho_- + 2\,\dot
P_-\,\Delta\sigma_\mathrm{H} = 0,
\end{displaymath}
\begin{displaymath}
\mu'_-\,\Delta\rho_+ + \mu'_+\,\Delta\rho_- +
2\,\dot\mu_-\,\Delta\sigma_\mathrm{H} = 0
\end{displaymath}
to lowest order.  Solving for $\Delta\rho_\pm$ yields the sensitivity of
the
symmetric and asymmetric shifts in the coexisting densities to a
change $\Delta\sigma_\mathrm{H}$ of the effective diameter as
\begin{displaymath}
\begin{array}{rl}
\displaystyle \frac{\Delta\rho_\pm}{\Delta\sigma_\mathrm{H}} 
&\displaystyle = 2 \, \frac{ P'_\pm\,\dot\mu_- - \mu'_\pm\,\dot P_-}{
P'_\mp\,\mu'_\pm -  P'_\pm\,\mu'_\mp}\\
\displaystyle &\displaystyle = \pm \frac{ P'_\pm\,\dot\mu_- -
\mu'_\pm\,\dot P_-}{ P'_l\,\mu'_v -  P'_v\,\mu'_l},
\end{array}
\end{displaymath}
where we have taken advantage of some cancellations in the denominator
to arrive at the second expression.

From the product of two density derivatives in the denominator as
opposed to single factors of density derivatives in the numerator it
is already apparent that the sensitivity of the binodal to variation
of $\sigma_\mathrm{H}$ must be proportional to the isothermal
compressibilities
$\kappa_v$ and $\kappa_l$ at the coexisting densities.  To make this more
explicit and simplify the expressions further, we introduce the free
energy per unit volume $f$ and use basic thermodynamic relations to
replace the pressure and chemical potential, as well as their
derivatives, by appropriate expressions in terms of $f$ and its
derivatives while converting reciprocals of $f''$ to the isothermal
compressibilities.  After some slightly tedious simplification we get
explicit expressions for the shifts in the coexisting densities
induced by a change $\Delta\sigma_\mathrm{H}$ of the effective diameter,
\textit{viz.{}},
\begin{displaymath}
\begin{array}{rl}
\displaystyle \frac{\Delta\rho_\pm}{\Delta\sigma_\mathrm{H}}
&\displaystyle =  \rho_l^2\,\kappa_l\,\left(\dot f'_l - \frac{\dot
f_-}{\Delta\rho_-}\right)
\pm
\rho_v^2\,\kappa_v\,\left(\dot f'_v - \frac{\dot
f_-}{\Delta\rho_-}\right)\\
\displaystyle &\displaystyle = \frac{\Delta\rho_l}{\Delta\sigma_\mathrm{H}}
\pm \frac{\Delta\rho_v}{\Delta\sigma_\mathrm{H}}.
\end{array}
\end{displaymath}
The change of any one of the coexisting densities with $\sigma_\mathrm{H}$
is thus
proportional to the product of the compressibility at that density and
a factor that combines the sensitivities of the free energy at both
branches of the binodal with its density derivative.

In an exact theory, of course, all the $\sigma_\mathrm{H}$ derivatives
vanish
exactly and the coexisting densities are strictly independent of
$\sigma_\mathrm{H}$ even at the critical point, $\beta= \beta_c$,
$\rho_-=0$, where
the compressibility diverges.  In an approximate theory, on the other hand,
$\dot
f' - \dot f_- / \Delta\rho_-$ is non-zero for $\beta> \beta_c$, and while
its
limit for $\beta\to \beta_c$ obviously vanishes, that of its product
with the diverging compressibility may be zero, finite, or even
infinite; for BAS\ SCOZA, Figs.~\ref{fig:coex:tr} and~\ref{fig:coex:or}
indicate a
finite value.  As $\beta$ further increases from $ \beta_c$, the two
branches of the binodal separate rapidly, and the one-sided difference
quotient $\dot f_- / \Delta\rho_-$ over the coexistence region no longer
cancels the derivative $\dot f'$ at either side of the binodal.  The
expression in parentheses is therefore no longer expected to be close
to zero.  In the case of BAS\ SCOZA, the variability of $f$ with
$\sigma_\mathrm{H}$
strongly increases with density, as is expected for thermodynamic
perturbation theories in general.  This carries over to $\dot f' -
\dot f_- / \Delta\rho_-$ and, in combination with the prefactor $\rho^2$,
so
explains why the sensitivity of $ \rho_l$ is significantly larger than
that of $ \rho_v$.

\newpage

\subsection*{Tables}

Tab.~\ref{tab:sensitivity}: Sensitivity of free energy, pressure, and
chemical potential at
$ \rho^* = 1$ and $ \rho^* = 0.7$ to the choice of $\sigma_\mathrm{H}$. 
For any
quantity $X$, $\Delta X$ is the difference between the results
obtained with $ a_\mathrm{BFC}=2.3$ and $ a_\mathrm{BFC}=2.5$.  The columns
$\Delta X/X$
and $\Delta X/(dX/d\rho)$ relate the response of both the value and
the slope of $X$ as a function of $\rho$ and so characterize the
sensitivity of the shape of $X(\rho)$ to variation of $ a_\mathrm{BFC}$. 
---
In general, the free energy is seen to be far more forgiving than
$P$ and $\mu$ with respect to the precise choice of $\sigma_\mathrm{H}$,
especially at low temperature.\par
\bigskip

\newpage

\subsection*{Figures}

\begin{figure}[p]\vbox{\noindent\includegraphics[width=\hsize              
,angle=-90]                    {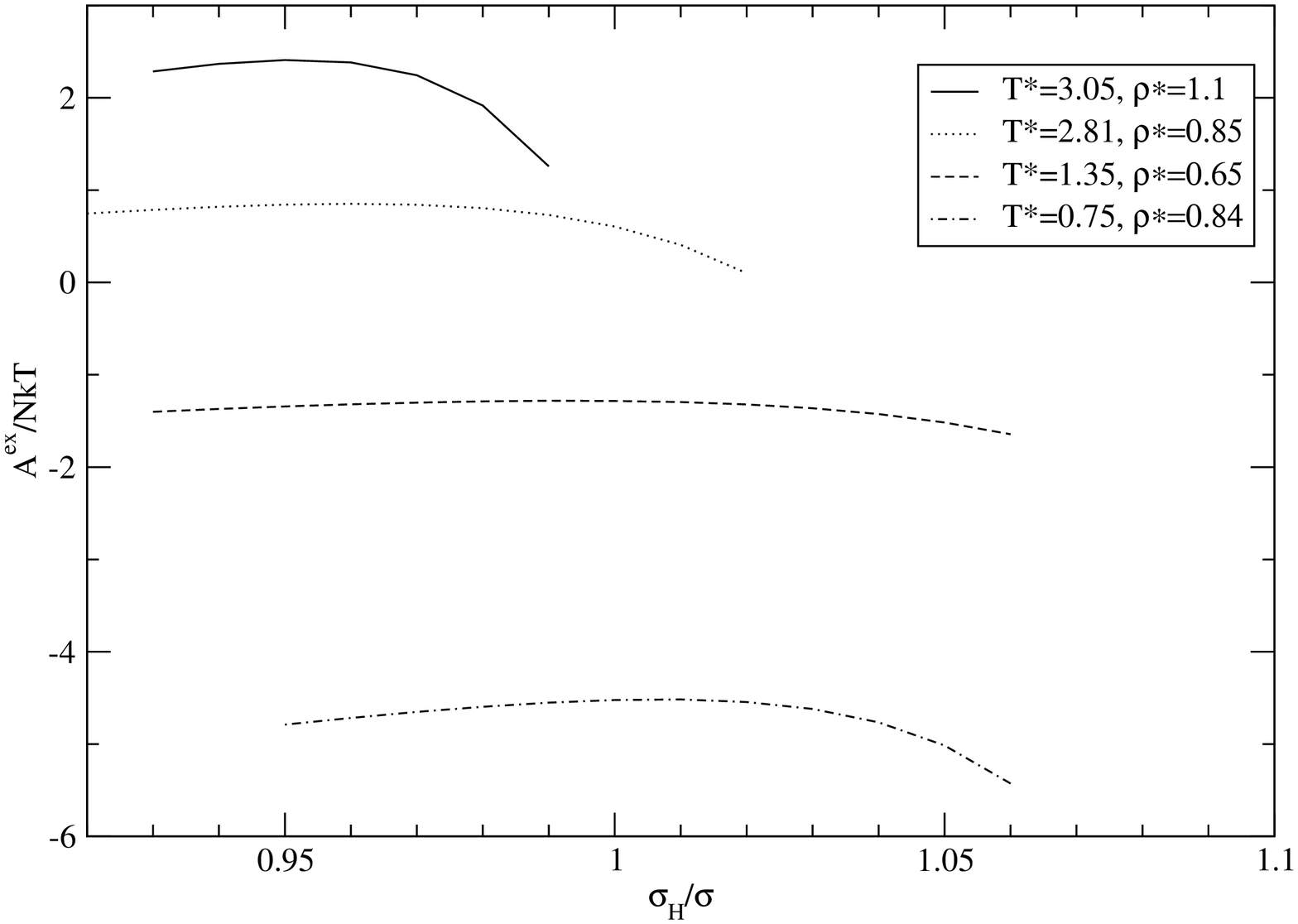}
\caption{}\label{fig:Aex}}
\end{figure}

Fig.~\ref{fig:Aex}: Dependence of the BAS\ SCOZA\ excess free energy $
A^\mathrm{ex}$ on the
effective hard core diameter $\sigma_\mathrm{H}$ for the four states
considered
in Ref.~\cite{bas:reformulation}.  (This corresponds to Fig.~2 of
Ref.~\cite{bas:reformulation}.)\par
\bigskip

\begin{figure}[p]\vbox{\noindent\includegraphics[width=\hsize              
,angle=-90]                    {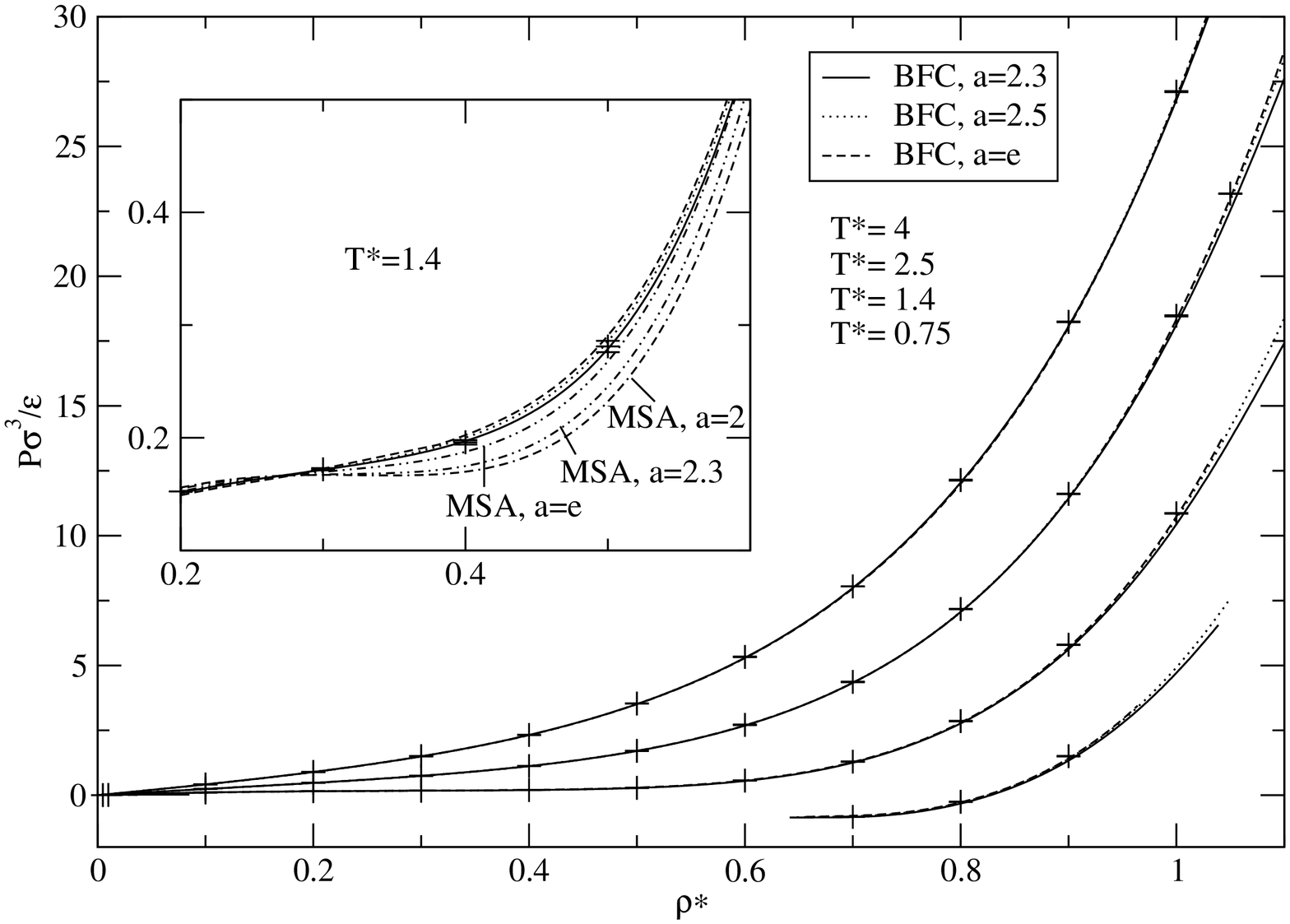}
\caption{}\label{fig:P}}
\end{figure}

Fig.~\ref{fig:P}: The pressure $P$ as a function of $\rho$ along four
isotherms as
computed within BAS\ SCOZA\ with the BFC\ diameters with $
a_\mathrm{BFC}=2.3$,
2.5 and $e$, as well as the simulation data of Ref.~\cite{gubbins}.
(This corresponds to Fig.~5 of Ref.~\cite{bas:analytical}.)  The inset
gives a detailed view of the isotherm at $ T^* = 1.4$, along with
the curves obtained in an MSA\ based variant of the theory
($K=\beta$) for $ a_\mathrm{BFC}=2$, 2.3, and $e$.\par
\bigskip

\begin{figure}[p]\vbox{\noindent\includegraphics[width=\hsize              
,angle=-90]                    {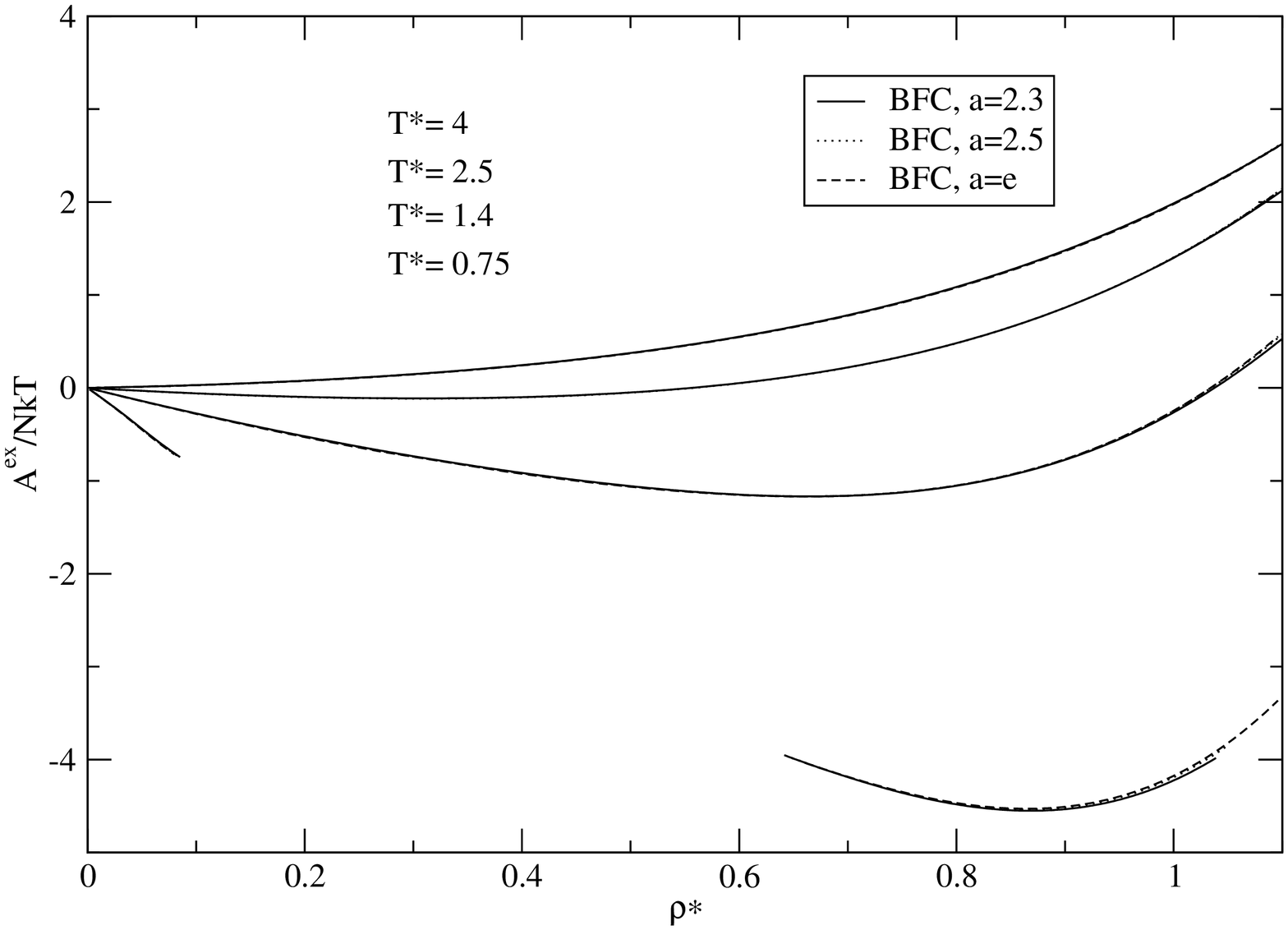}
\caption{}\label{fig:A:ex}}
\end{figure}

Fig.~\ref{fig:A:ex}: The excess free energy $ A^\mathrm{ex}$ as a function
of $\rho$ along four
isotherms as computed with the BFC\ diameter with $ a_\mathrm{BFC}=2.3$,
2.5, and $e$.\par
\bigskip

\begin{figure}[p]\vbox{\noindent\includegraphics[width=\hsize              
,angle=-90]                    {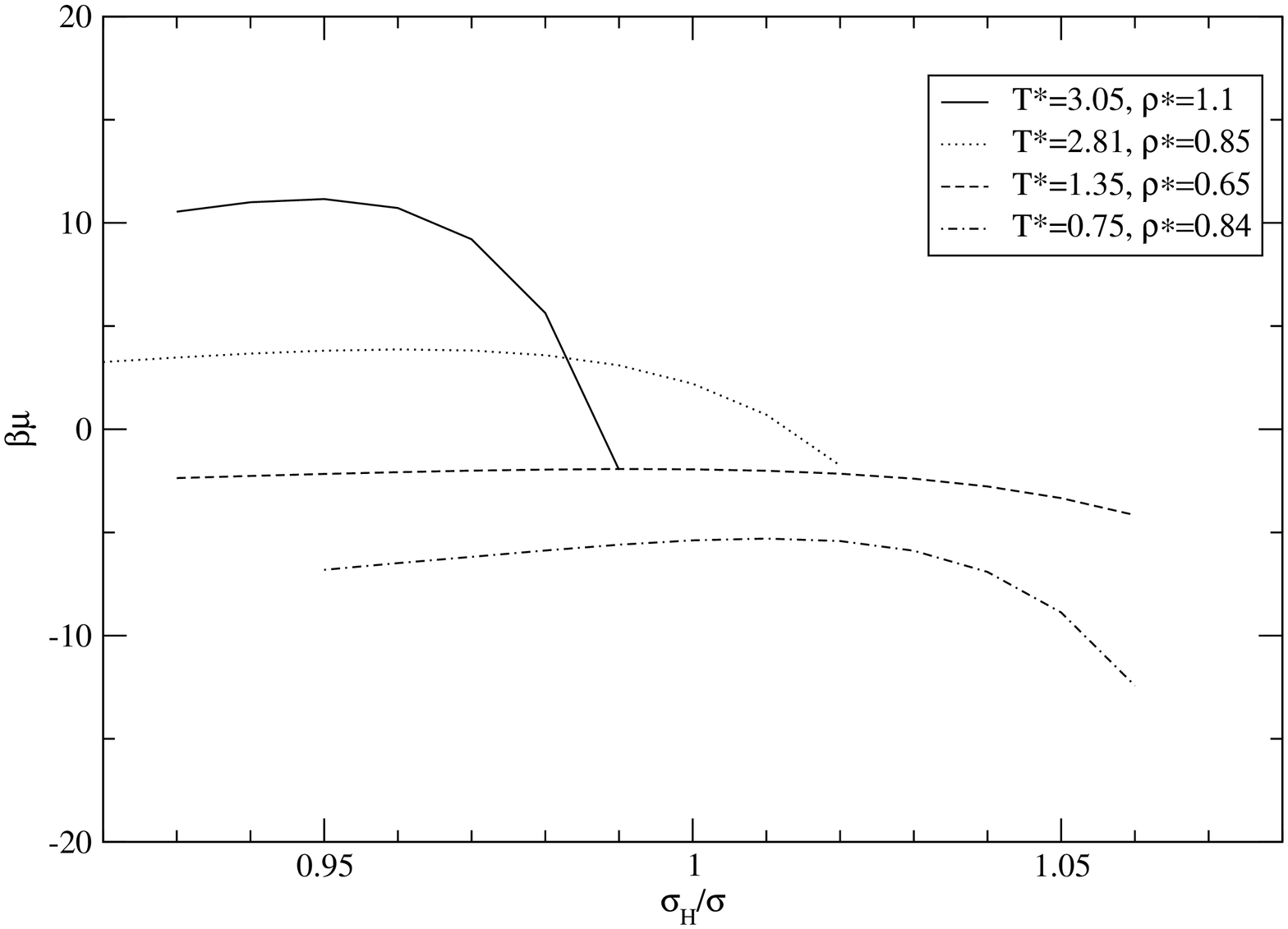}
\caption{}\label{fig:mu}}
\end{figure}

Fig.~\ref{fig:mu}: Dependence of the chemical potential $\mu$ on the
effective hard
core diameter $\sigma_\mathrm{H}$ for the four states considered in
Ref.~\cite{bas:reformulation}.\par
\bigskip

\begin{figure}[p]\vbox{\noindent\includegraphics[width=\hsize              
,angle=-90]                    {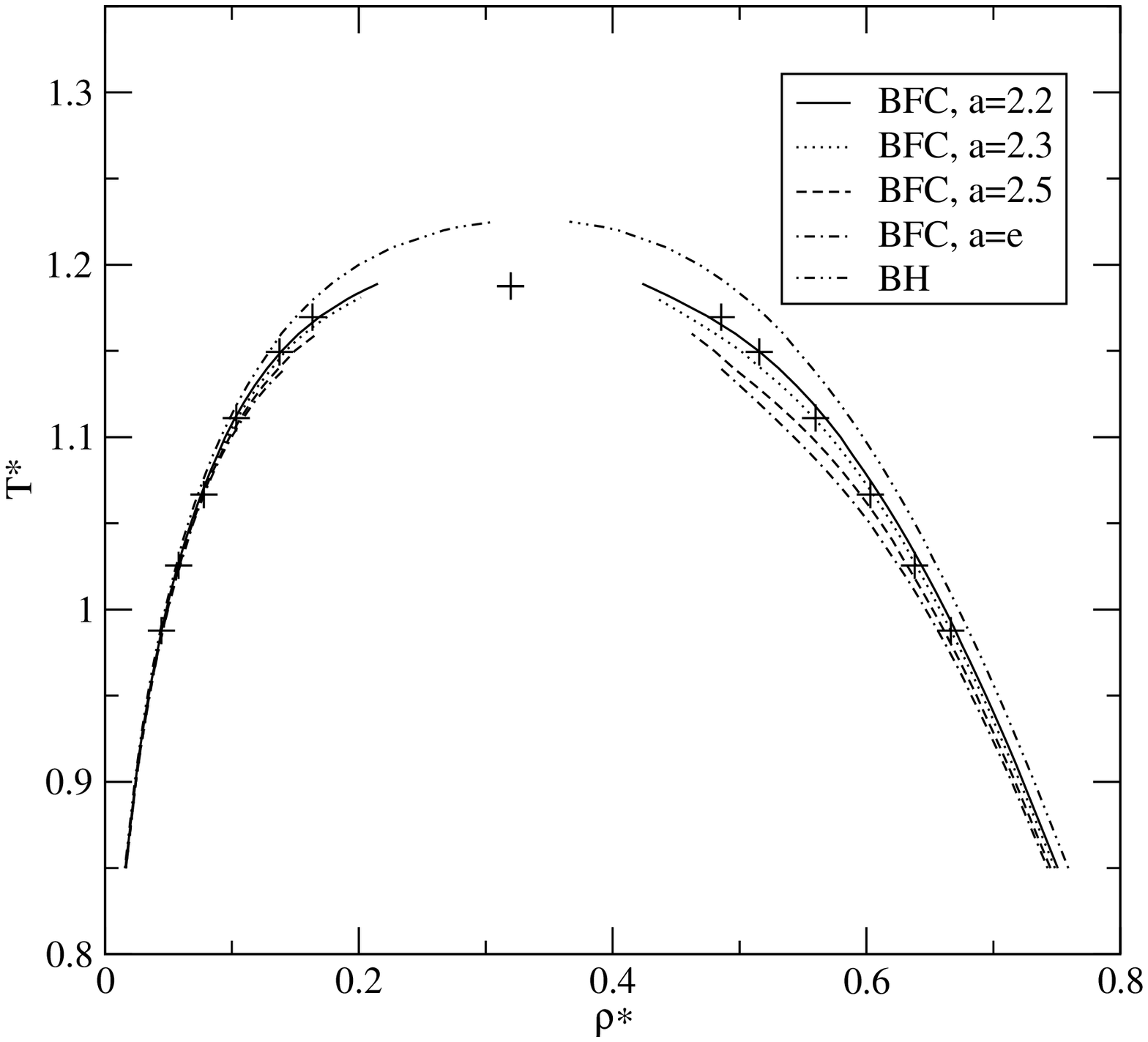}
\caption{}\label{fig:coex:tr}}
\end{figure}

Fig.~\ref{fig:coex:tr}: The curve of phase coexistence as obtained for $
a_\mathrm{BFC}=2.2$, 2.3,
2.5, and $e$ as well as with the BH\ diameter, along with
simulation results of Ref.~\cite{wilding}.  All data refer to a truncated
but not shifted, rather than the full, LJ\ potential.\par
\bigskip

\begin{figure}[p]\vbox{\noindent\includegraphics[width=\hsize              
,angle=-90]                    {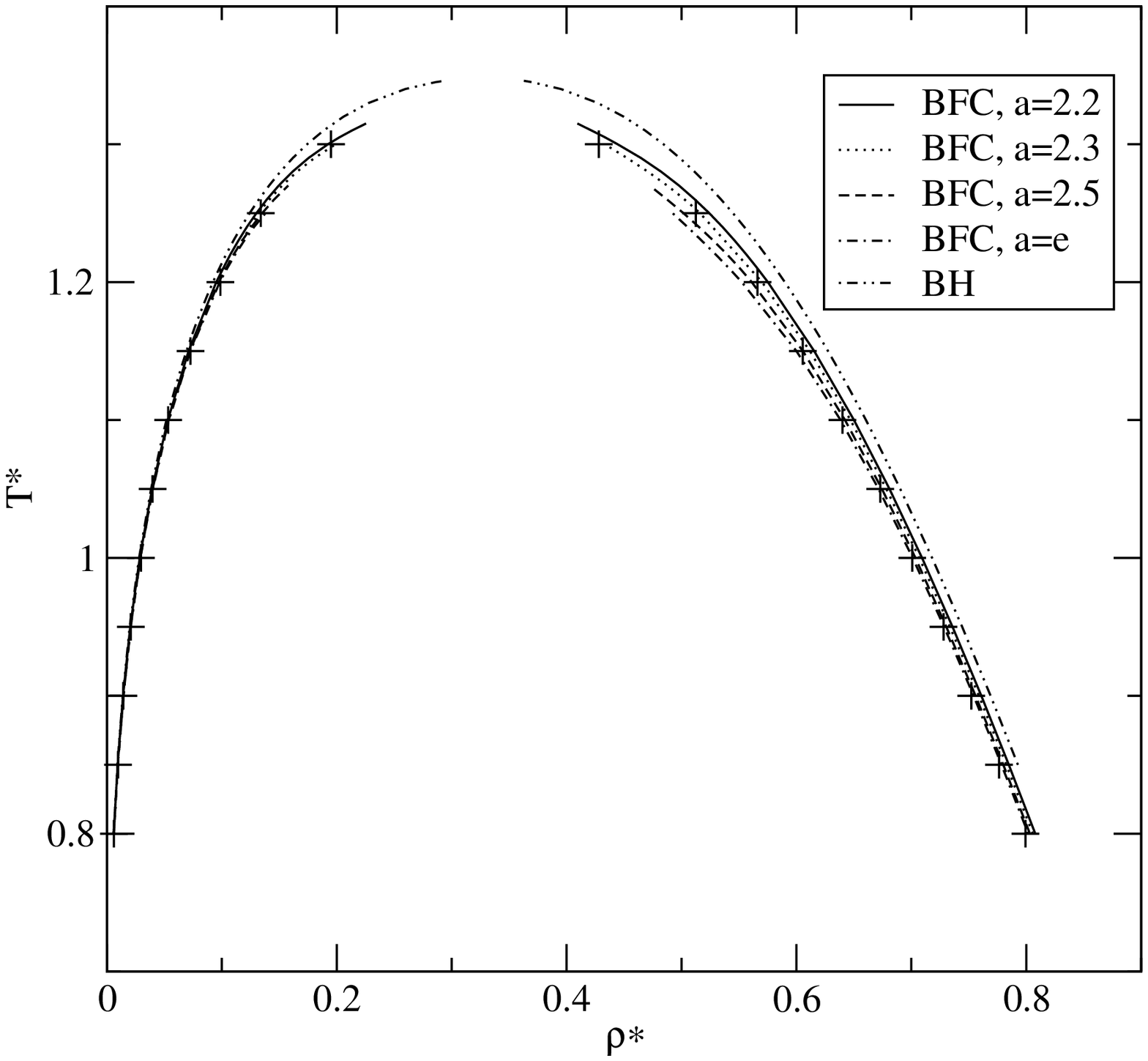}
\caption{}\label{fig:coex:or}}
\end{figure}

Fig.~\ref{fig:coex:or}: The curve of phase coexistence as obtained for $
a_\mathrm{BFC}=2.2$, 2.3,
2.5, and $e$ as well as with the BH\ diameter, along with
simulation results of Ref.~\cite{lj:3}.  All data refer to the full LJ\
potential.\par
\bigskip

\begin{figure}[p]\vbox{\noindent\includegraphics[width=\hsize              
,angle=-90]                    {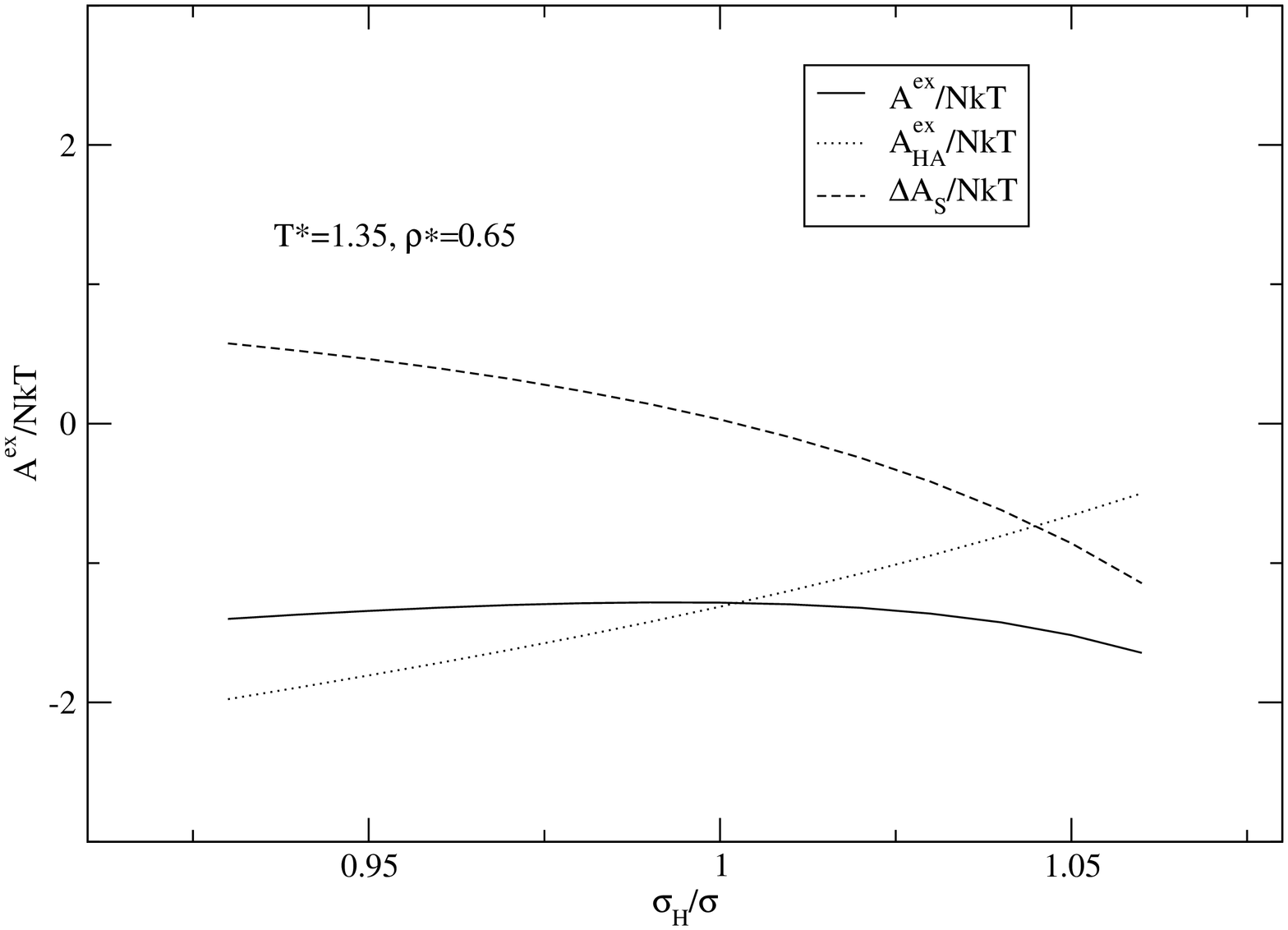}
\caption{}\label{fig:deltaA}}
\end{figure}

Fig.~\ref{fig:deltaA}: SCOZA\ ($ A^\mathrm{ex}_\mathrm{HA}$) and BAS\
($\Delta A_\mathrm{S}$) contributions to the BAS\ SCOZA\
excess free energy at $ T^* = 1.35$, $ \rho^* = 0.65$ for varying
$\sigma_\mathrm{H}$.  Clearly, the SA\ free energy $ A^\mathrm{ex} = 
A^\mathrm{ex}_\mathrm{HA} + \Delta A_\mathrm{S}$ is far
less sensitive to $\sigma_\mathrm{H}$ than either $
A^\mathrm{ex}_\mathrm{HA}$ or $\Delta A_\mathrm{S}$ taken by
itself.  On the other hand, $ A^\mathrm{ex}$ is not quite constant as must
be the case for
an exact theory.\par
\bigskip

\begin{figure}[p]\vbox{\noindent\includegraphics[width=\hsize              
,angle=-90]                    {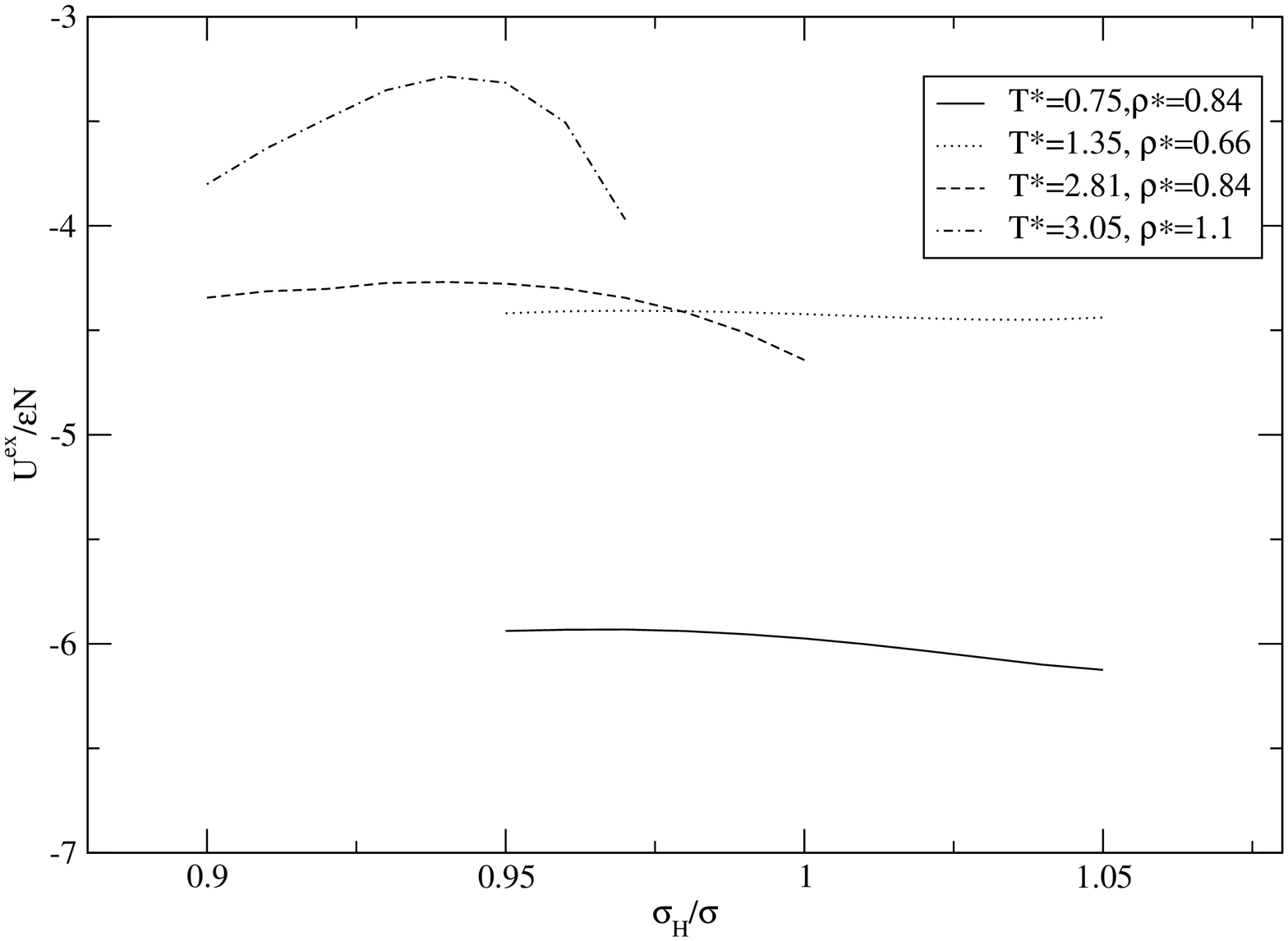}
\caption{}\label{fig:Uex}}
\end{figure}

Fig.~\ref{fig:Uex}: Diameter dependence of the internal energy at four
different
states.\par
\bigskip

\begin{figure}[p]\vbox{\noindent\includegraphics[width=\hsize              
]                    {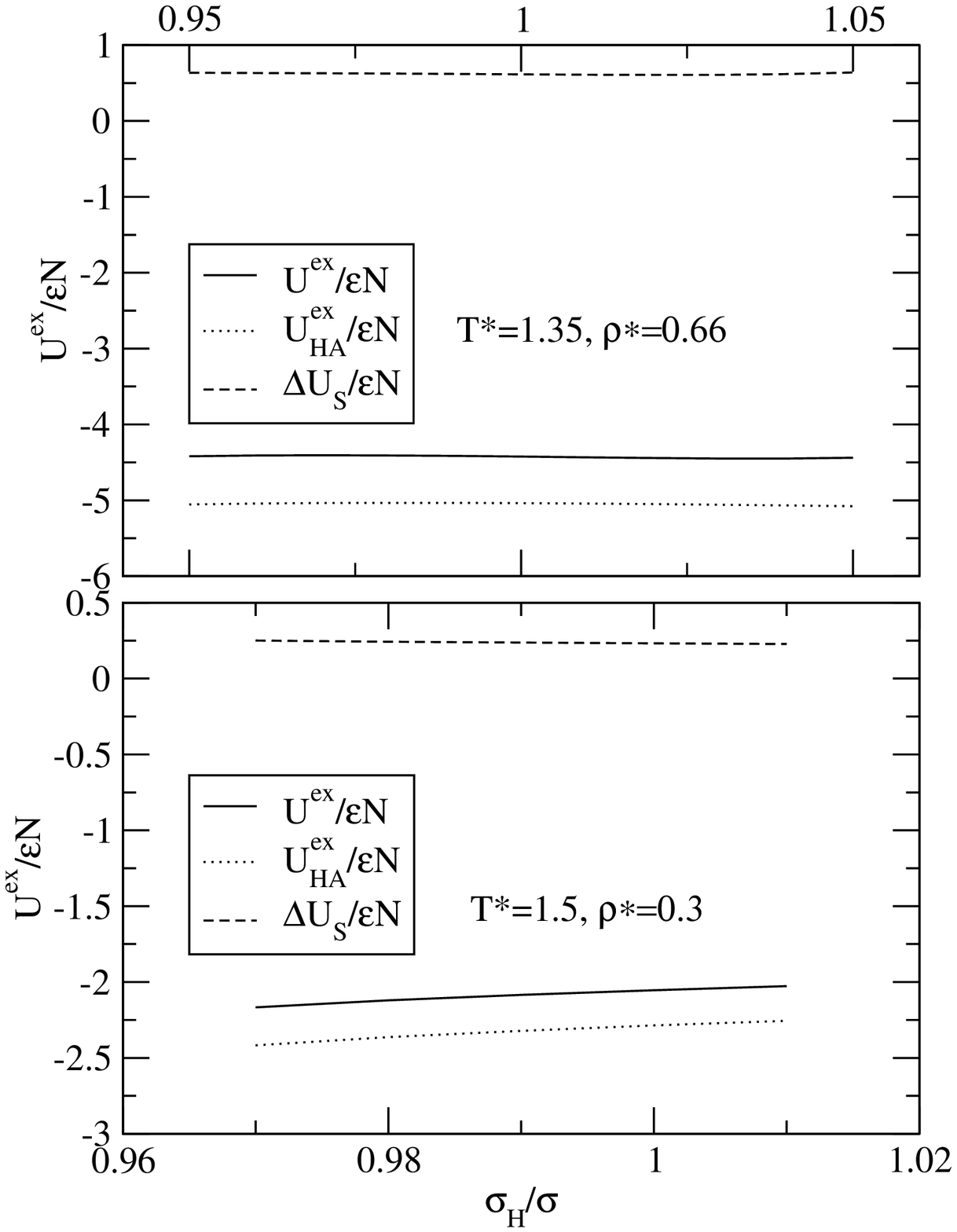}
\caption{}\label{fig:U:split}}
\end{figure}

Fig.~\ref{fig:U:split}: SCOZA\ ($ U^\mathrm{ex}_\mathrm{HA}$) and BAS\
($\Delta U_\mathrm{S}$) contributions to the BAS\ SCOZA\
internal energy at $ T^* = 1.35$, $ \rho^* = 0.65$ and $ T^* = 1.5$,
$ \rho^* = 0.3$ for varying $\sigma_\mathrm{H}$.\par
\bigskip

\begin{figure}[p]\vbox{\noindent\includegraphics[width=\hsize              
,angle=-90]                    {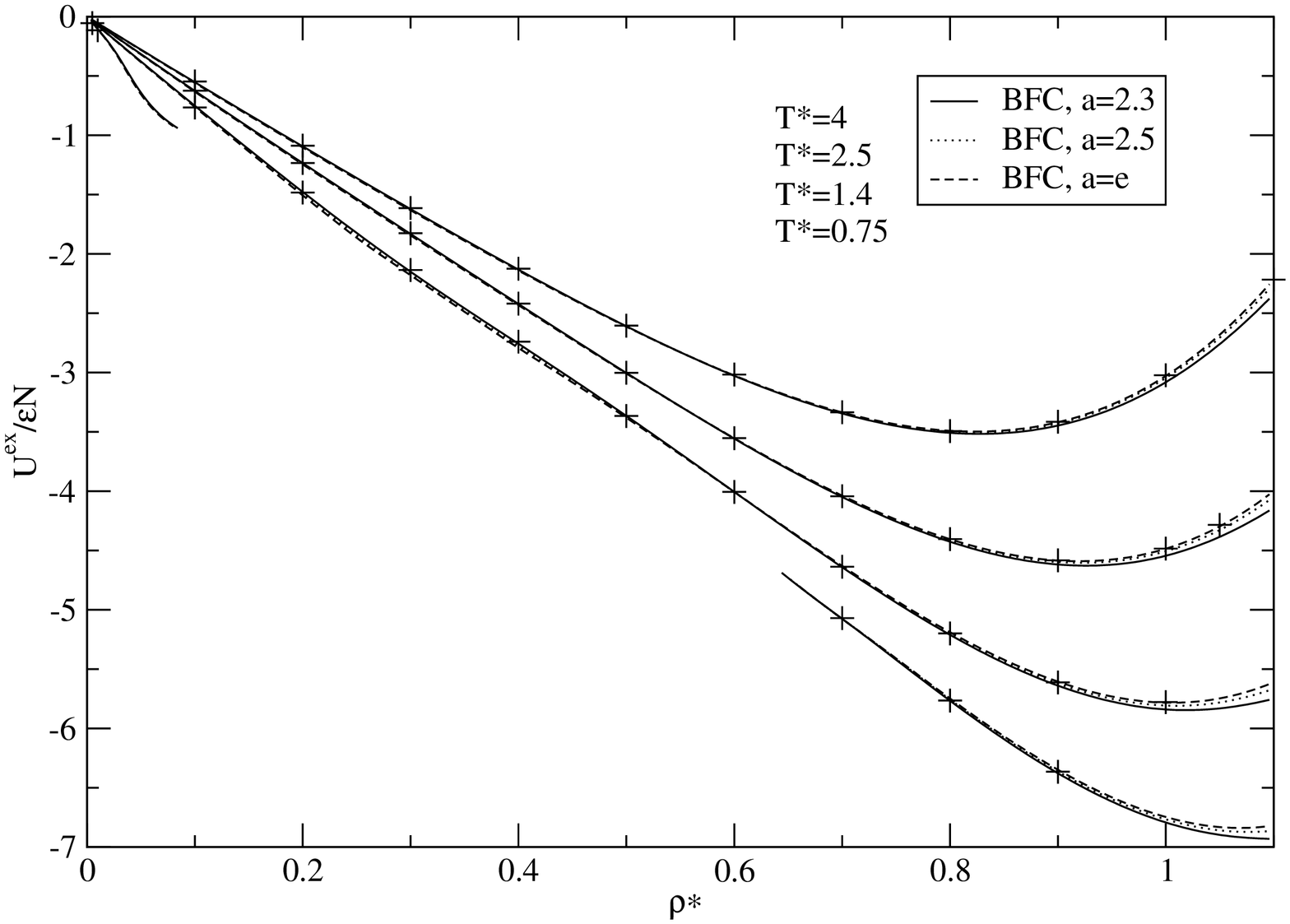}
\caption{}\label{fig:U}}
\end{figure}

Fig.~\ref{fig:U}: BAS\ SCOZA\ and simulation \cite{gubbins} results for the
configurational
internal energy $ U^\mathrm{ex}$ along four different isotherms for three
choices of $ a_\mathrm{BFC}$.  (This corresponds to one panel of Fig.~4 of
Ref.~\cite{bas:reformulation}.)\par
\bigskip

\newpage

\newpage\begin{table}
\begin{tabular}{|ll|cc|cc|cc|}
$\rho^*$&$T^*$&$\left\vert\frac{\Delta
A^\mathrm{ex}}{ A^\mathrm{ex}}\right\vert$&$\left\vert\frac{\Delta
A^\mathrm{ex}}{d
A^\mathrm{ex}/d\rho^*}\right\vert$&$\left\vert\frac{\Delta
P}P\right\vert$&$\left\vert\frac{\Delta
P}{dP/d\rho^*}\right\vert$&$\left\vert\frac{\Delta
\mu}\mu\right\vert$&$\left\vert\frac{\Delta
\mu}{d\mu/d\rho^*}\right\vert$\\
\hline
0.7&
4.0
 &$2.6\cdot10^{-3}$&$7.8\cdot10^{-4}$&$8.0\cdot10^{-4}$&$2.0\cdot10^{-4}$&$9.5\cdot10^{-3}$&$1.7\cdot10^{-3}$\\&
2.5
 &$5.0\cdot10^{-3}$&$5.2\cdot10^{-4}$&$1.1\cdot10^{-3}$&$2.3\cdot10^{-4}$&$1.0\cdot10^{-2}$&$1.0\cdot10^{-3}$\\&
1.4
 &$8.5\cdot10^{-4}$&$2.5\cdot10^{-3}$&$1.2\cdot10^{-2}$&$1.4\cdot10^{-3}$&$2.6\cdot10^{-3}$&$3.1\cdot10^{-4}$\\&
 
 0.75&$6.7\cdot10^{-4}$&$7.6\cdot10^{-4}$&$3.5\cdot10^{-2}$&$2.6\cdot10^{-2}$&$6.6\cdot10^{-3}$&$2.3\cdot10^{-2}$\\
\hline
1.0&
4.0
 &$4.0\cdot10^{-4}$&$1.4\cdot10^{-4}$&$4.8\cdot10^{-3}$&$1.2\cdot10^{-3}$&$2.1\cdot10^{-3}$&$5.9\cdot10^{-4}$\\&
2.5
 &$3.2\cdot10^{-3}$&$7.1\cdot10^{-4}$&$1.0\cdot10^{-2}$&$2.2\cdot10^{-3}$&$8.3\cdot10^{-3}$&$1.9\cdot10^{-3}$\\&
1.4
 &$6.2\cdot10^{-2}$&$2.4\cdot10^{-3}$&$2.1\cdot10^{-2}$&$3.5\cdot10^{-3}$&$2.5\cdot10^{-2}$&$3.6\cdot10^{-3}$\\&
 
 0.75&$8.0\cdot10^{-3}$&$6.2\cdot10^{-3}$&$4.4\cdot10^{-2}$&$4.6\cdot10^{-3}$&$2.4\cdot10^{-1}$&$5.0\cdot10^{-3}$\\
\end{tabular}
\caption{}\label{tab:sensitivity}
\end{table}

\newpage

\newpage

\newpage

\newpage

\newpage

\newpage

\newpage

\newpage

\newpage

\newpage


\newpage

\begin{thebibliography}{99}

\bibitem{scoza:12}J.~S.\ H\o ye, G. Stell, J.\ Chem.\ Phys.\ {\bf 67},
439 (1977).
\bibitem{scoza:11}J.~S.\ H\o ye, G.\ Stell, Mol.\ Phys.\ {\bf 52},
1071 (1984).
\bibitem{scoza:9}J.~S.\ H\o ye, G.\ Stell, Int.\
J.\ Thermophys.\ {\bf 6}, 561 (1985).
\bibitem{scoza:1}R.\ Dickman, G.\ Stell, Phys.\ Rev.\ Lett.\ {\bf 77},
996 (1996).
\bibitem{scoza:8}D.\ Pini, G.\ Stell, R.\ Dickman, Phys.\ Rev.\ E\ {\bf
57}, 2862 (1998).
\bibitem{scoza:6}D.\ Pini, G.\ Stell, N.~B.\
Wilding, Mol.\ Phys.\ {\bf 95}, 483 (1998).
\bibitem{plwk:bin:symm}E.\ Sch\"oll-Paschinger, D.\
Levesque, J.-J.\ Weis, G.\ Kahl, J.\ Chem.\ Phys.\ {\bf 122}, 024507
(2005).
\bibitem{scoza:25}J.\ S.\ H\o ye, D.\ Pini, G.\
Stell, Physica A\ {\bf 279}, 213 (2000).
\bibitem{scoza:15}D.\ Pini, G.\ Stell, N.\ B.\
Wilding, J.\ Chem.\ Phys.\ {\bf 115}, 2702 (2001).
\bibitem{paschinger:dr}E.\ Sch\"oll-Paschinger, {\it Phase
behavior of simple fluids and their mixtures}, PhD\ thesis, Technische
Universit\"at Wien (2002).
\bibitem{scoza:28}E.\ Sch\"oll-Paschinger, J.\ Chem.\ Phys.\ {\bf 120},
11698(2004).
\bibitem{scoza:26}E.\ Sch\"oll-Paschinger, A.\ L.\
Benavides, R.\ Casta\~neda-Priego, J.\ Chem.\ Phys.\ {\bf 123}, 234513
(2005).
\bibitem{scoza:27}B.~M.\
Mladek, G.\ Kahl, M.\ Neumann, J.\
Chem.\ Phys.\ {\bf 124}, 064503 (2006).
\bibitem{ar:21}J.\ S.\
H\o ye, A.~Reiner, accepted for publication in J.\ Chem.\ Phys.
\bibitem{bas:analytical}D.~Ben-Amotz, G.~Stell, J.\ Chem.\ Phys.\ {\bf
119}, 10777 (2003).  Erratum:

J.\ Chem.\ Phys.\ {\bf 120},
4994 (2004).
\bibitem{bas:reformulation}D.~Ben-Amotz, G.~Stell, J.\ Phys.\ Chem.\ B\
{\bf 108}, 6877 (2004).
\bibitem{wca}J.\ D.\ Weeks, D.\ Chandler, H.\ C.\
Andersen, J.\ Chem.\ Phys.\ {\bf 54}, 5237 (1971).
\bibitem{tang}Y.~Tang, J.\ Chem.\ Phys.\ {\bf 116}, 6694 (2002).
\bibitem{barker:henderson}J.\ A.\ Barker, D.\
Henderson, J.\ Chem.\ Phys.\ {\bf 47}, 4714 (1967).
\bibitem{allg:27}F.\ O.\ Raineri, G.\ Stell, D.\
Ben-Amotz, J.\ Phys.: Condens.\ Matter\ {\bf 16}, S4887 (2004).
\bibitem{bas:hard}D.~Ben-Amotz, G.~Stell, J.\ Chem.\ Phys.\ {\bf 120},
4844 (2004).
\bibitem{hs:2}D.\ Henderson, E.~W.\ Grundke, J.\ Chem.\ Phys.\ {\bf 63},
601 (1975).
\bibitem{waisman}E.\ Waisman, Mol.\ Phys.\ {\bf 32}, 1627 (1973).
\bibitem{hansen:mcdonald}J.~P.\ Hansen, I.~R.\
McDonald, {\it Theory of simple liquids}, London (Academic)
${}^21986$.
\bibitem{gubbins}J.\ K.\ Johnson, J.\ A.\ Zollweg,
K.\ E.\ Gubbins, Mol.\ Phys.\ {\bf 78}, 591 (1993).
\bibitem{lj:3}A.\ Lotfi, J.\ Vrabec, J.\ Fischer, Mol.\ Phys.\ {\bf 76},
1319 (1992).
\bibitem{wilding}N.\ B.\ Wilding, Phys.\ Rev.\ E\ {\bf 52}, 602
(1995).
\bibitem{lj:2}J.\ J.\ Potoff, A.\ Z.\
Panagiotopoulos, J.\ Chem.\ Phys.\ {\bf 109}, 10914 (1998).
\bibitem{caillol}J.\ M.\ Caillol, J.\ Chem.\
Phys.\ {\bf 109}, 4885 (1998).
\bibitem{smit}B.\ Smit, J.\ Chem.\ Phys.\ {\bf 96}, 8639 (1992).

\end{thebibliography}
\end{document}